\documentclass[10pt]{IEEEtran}

\usepackage{cite}
\usepackage{amsmath,amssymb,amsfonts}
\usepackage{algorithmic}
\usepackage{graphicx}
\usepackage{textcomp}
\usepackage{subfigure}
\usepackage{authblk}
\def\BibTeX{{\rm B\kern-.05em{\sc i\kern-.025em b}\kern-.08em
		T\kern-.1667em\lower.7ex\hbox{E}\kern-.125emX}}
\begin{document}
	\title{mmWave/THz Channel Estimation Using Frequency-Selective Atomic Norm Minimization}
    \author{Yicheng Xu, Hongyun Chu, Xiaodong Wang}

	
	
	
	
	
	\maketitle
	
	\begin{abstract}
		We propose a MIMO channel estimation method for millimeter-wave (mmWave) and terahertz (THz) systems based on frequency-selective atomic norm minimization (FS-ANM). For the strong line-of-sight property of the channel in such high-frequency bands, prior knowledge on the ranges of angles of departure/arrival (AoD/AoA) can be obtained as the prior knowledge, which can be exploited by the proposed channel estimator to improve the estimation accuracy. Simulation results show that the proposed method can achieve considerable performance gain when compared with the existing approaches without incorporating the the strong line-of-sight property.
	\end{abstract}
	
	
	\section{Introduction}
	
	\let\thefootnote\relax\footnote{Yicheng Xu is with the National Mobile Communications Research Laboratory, Southeast University, Nanjing 210096, China (e-mail: 421570704@qq.com). \textit{(Corresponding author: Yicheng Xu.)}}\let\thefootnote\relax\footnote{Hongyun Chu is with the School of Communications and Information Engineering, Xi'an University of Posts and Telecommunications, Xi'an, 710121, China (e-mail: hy\_chu@foxmail.com).}\let\thefootnote\relax\footnote{Xiaodong Wang is with the Department of Electrical Engineering, Columbia University, New York 10027, USA (e-mail: wangx@ee.columbia.edu).}The mmWave/THz communication has been deemed as a potential solution for future wireless communication systems \cite{c01,c02}. To compensate for the severe signal propagation loss at mmWave/THz band, the systems are expected to configure with ultra-massive antenna arrays at
	transceivers to achieve sufficient beamforming gains. For such MIMO systems, it is well known that the channel state information (CSI) is indispensable for reliable signal transmission and reception, and especially useful for designing efficient beamformers in mmWave/THz band \cite{c13}. However, channel estimation is challenging in mmWave/THz systems where the number of antennas is large and the received signal-to-noise ratio (SNR) is low. Most of the existing estimation methods are based on the rich scattering assumption of the channel, which limits their applications due to high training overhead and computational cost.
	
	The inherent sparsity \cite{c14} of the ultra-massive MIMO channel has been used for reducing the training overhead and/or improving the estimation accuracy \cite{c15,c16,c17,c18,c19,c20,c21,c22,c23,c24,c25}, 
	which are overall classified into the closed-loop methods, e.g., \cite{c15,c16}, and the open-loop methods, e.g., \cite{c17,c18,c19,c20,c21,c22,c23,c24,c25}, both using grid-based compressive sensing (CS). To be specific, the closed-loop 
	schemes estimate the channels via the multistage beam search operation, and their performance is limited by the resolution of the pre-determined codebook, i.e., the better performance with the cost of larger storage, more complex parsing process and longer time delay; the open-loop techniques perform channel estimation without the feedback process by using the pilot-based multiple signal classification (MUSIC) methods. In particular, these on-grid methods suffer from the basis mismatch problem.
    In view of the continuous valued angles of channel, a gridless approach \cite{c26}, which uses atomic norm minimization (ANM) to manifest the signal sparsity in the continuous parameter domain, has been proposed for 
    angular estimation \cite{c29,c30}.
    Under certain conditions, ANM based denoising methods can achieve exact sparse signals reconstruction, avoiding the effects of basis mismatch which can plague grid-based CS techniques. 
	
	In mmWave/THz systems, the channels exhibit strong line-of-sight. Hence it is possible to obtain prior knowledge on the ranges of AoD/AoAs which can push the solution to the reduced feasible regions. In this paper, we propose a channel estimator that can incorporate such prior knowledge, based on frequency-selective atomic norm minimization. 

	\section{System Model}
	
	We consider a downlink ultra-massive MIMO communication system working at mmWave/THz band, where a base station (BS) equipped with $N_t$ antennas transmits data to a user equipment (UE) equipped with $N_r$ antennas. The BS-UE channel $\mathbf{H}$ can be expressed as\cite{c32}
	\begin{equation}
	\mathbf{H}=\sum\nolimits_{l=1}^{L}\alpha_l\mathbf{a}(N_r,\phi_l)\mathbf{a}^H(N_t,\theta_l),\label{eq:H}
	\end{equation}
	where $\alpha_l\sim\mathcal{CN}(0,\bar{P}/\sqrt{\rho})$ is the complex gain of the $l$th path, $l=1,...,L$, with $\bar{P}$ and $\rho$ denoting the average power gain and the average path loss between the BS and the UE respectively. $\mathbf{a}(N_t,\theta_l)$ and $\mathbf{a}(N_r,\phi_l)$ denote the antenna array response vectors of the BS and the UE respectively. In this paper, we consider the uniform linear arrays (ULA), where array response is in the form of
	\begin{equation}
	\mathbf{a}(N,\phi)=[1,e^{i2\pi \phi},...,e^{i2\pi(N-1)\phi}]^T,
	\end{equation}
	In \eqref{eq:H}, $\phi_l=(d/\lambda )\sin (\bar{\phi}_l)$ and $\theta_l=(d/\lambda )\sin (\bar{\theta}_l)$, with $\lambda$ denoting the signal wavelength, $d$ denoting the interval between adjacent antenna elements, and $\bar{\phi}_l,\bar{\theta}_l$ being the UE's azimuth AoA and the BS' azimuth AoD of the $l$th path respectively. 
	
	In this paper, we assume that the ranges of the AoD/AoAs are known a priori, i.e., $\forall l$, $\bar{\theta}_l\in\Omega_1$, $\bar{\phi}_l\in\Omega_2$ with $\Omega_1,\Omega_2\subset [0,2\pi]$. Thus we have $\theta_l\in\mathcal{I}_1$ and $\phi_l\in\mathcal{I}_2$, with $\mathcal{I}_1,\mathcal{I}_2\subset[-d/\lambda,d/\lambda]$, $l=1,...,L$. Without loss of generality, we set $d/\lambda=1/2$ in this paper.

	
	The channel $\mathbf{H}$ in \eqref{eq:H} can be rewritten in the matrix form as
	\begin{equation}
	\mathbf{H}=\mathbf{A}_r\boldsymbol{\Lambda}\mathbf{A}_t^H,\theta_l\in\mathcal{I}_1,\phi_l\in\mathcal{I}_2,\label{eq:H1}
	\end{equation}
	where $\boldsymbol{\Lambda}=\mathrm{diag}(\alpha_1,...,\alpha_L)$, and the matrices $\mathbf{A}_t=[\mathbf{a}(N_t,\theta_1),...,\mathbf{a}(N_t,\theta_L)]$ and $\mathbf{A}_r=[\mathbf{a}(N_r,\phi_1),...,\mathbf{a}(N_r,\phi_L)]$ contain the array response of the BS and the UE respectively. 
	
	
	To estimate the channel matrix, the transmitter sends $S$ distinct beams during $S$ successive time slots, i.e., in the $s$-th time slot, the beamforming vector $\mathbf{f}_s\in\mathbb{C}^{N_t}$ is selected from a codebook\cite{c34}. Thus the received signal of the $s$-th time slot can be expressed as
	\begin{equation}
	\mathbf{y}_s=\mathbf{H}\mathbf{f}_sx_s+\mathbf{n}_s,\label{eq:y}
	\end{equation}
	where $\mathbf{n}_s\sim\mathcal{CN}(\mathbf{0},\sigma_n^2\mathbf{I}_{N_r})$ is the additive white Gaussian noise with $\mathbf{I}_{N_r}$ denoting the $N_r\times N_r$ identity matrix, and $x_s$ denotes the pilot symbol in the $s$-th time slot. After the receiver collecting $\mathbf{y}_s\in\mathbb{C}^{N_r}$ for $s=1,...,S$, the obtained signal matrix
	\begin{equation}
	\mathbf{Y}=[\mathbf{y}_1,...,\mathbf{y}_S]=\mathbf{H}\mathbf{F}\mathbf{X}+\mathbf{N},\label{eq:Y}
	\end{equation}
	where $\mathbf{F}=[\mathbf{f}_1,...,\mathbf{f}_S]\in \mathbb{C}^{N_t\times S}$ consists of the beamforming vectors of the $S$ time slots, $\mathbf{X}=\mathrm{diag}(x_1,...,x_S)\in \mathbb{C}^{S\times S}$, and $\mathbf{N}=[\mathbf{n}_1,...,\mathbf{n}_S]\in \mathbb{C}^{N_r\times S}$. Our aim is to estimate $\mathbf{H}$ from $\mathbf{Y}$.
	
	\section{Channel estimation using FS-ANM}
	In this section, we first present the channel estimator for the case that the UE has only one antenna, and then for the case that the UE has multiple antennas.
	
	\subsection{Single Rx Antenna}
	
	When the UE has only one antenna, i.e., $N_t>1,N_r=1$, then $y_s$ is a scalar. We denote
	\begin{equation}
	\tilde{\mathbf{y}}=[y_1,...,y_S]^H=\mathbf{X}^H\mathbf{F}^H\tilde{\mathbf{h}}+\tilde{\mathbf{n}},\label{eq:y_1d}
	\end{equation}
	where $\tilde{\mathbf{h}}=\mathbf{H}^H|_{N_r=1}=\sum\nolimits_{l=1}^L\alpha_l^*\mathbf{a}(N_t,\theta_l)$, $\tilde{\mathbf{n}}=[n_1,...,n_S]^H$.
	
	To solve the off-grid problem, we employ the FS atomic norm to enforce the sparsity of $\tilde{\mathbf{h}}$. First, we briefly introduce the concept of FS Vandermonde decomposition and FS atomic norm\cite{c33}. 
	
	Define $\mathcal{I}=(f_L,f_H)\subset[-\frac{1}{2},\frac{1}{2}]$ as a frequency interval, and trigonometric polynomial 
	\begin{equation}
	\beta(f)=r_1z^{-1}+r_0+r_{-1}z, \label{eq:beta}
	\end{equation}
	where $z=e^{i2\pi f}$, $r_1=e^{i\pi(f_L+f_H)}\mathrm{sgn}(f_H-f_L)$, $r_0=-2\cos(\pi(f_H-f_L))\mathrm{sgn}(f_H-f_L)$, $r_{-1}=r_1^*$. Then $\beta(f)$ is always positive for $f\in\mathcal{I}$, and negative for $f\in[-\frac{1}{2},\frac{1}{2}]\setminus\mathcal{I}$.
	
	Given $\mathcal{I}\subset[-\frac{1}{2},\frac{1}{2}]$, a Toeplitz matrix $\mathbf{T}\in\mathbb{C}^{N\times N}$ with $r=\mathrm{rank}(\mathbf{T})\leq N-1$ admits a unique FS Vandermonde decomposition as $\mathbf{T}=\sum\nolimits_{k=1}^rc_k\mathbf{a}(N,f_k)\mathbf{a}^H(N,f_k)$ with $f_k\in\mathcal{I}$, if and only if 
	\begin{equation}\left\{\begin{split}
	&\mathbf{T}\succeq 0\\
	&\mathbf{T}_{\beta}\succeq 0
	\end{split}\right. ,\label{eq:tb}
	\end{equation}
	where $\mathbf{T}=\mathrm{Toep}(\mathbf{t})$ is generated by a complex sequence $\mathbf{t}=[t_{-N+1},t_{-N+2},...,t_{N-1}]^T$, where $\mathrm{Toep}(\cdot)$ denotes the Toeplitz matrix whose first column is the last $N$ elements of the input vector, $c_k>0$, $\mathbf{T}_{\beta}$ is a Toeplitz matrix defined as $[\mathbf{T}_{\beta}]_{mn}=\sum\nolimits_{j=-1}^1 r_jt_{m-n-j}$, $1\leq m,n\leq N-1$.

	From \eqref{eq:y_1d}, the class of signals is $\tilde{\mathbf{h}}=\sum\nolimits_{l=1}^L\alpha_l^*\mathbf{a}(N_t,\theta_l)$. Therefore the FS atom is of the form $\mathbf{a}(N_t,\theta)\in \mathbb{C}^{N_t\times 1}$. The FS atomic set is defined as $\mathcal{A}_{\mathcal{I}}=\{\mathbf{a}(N_t,\theta)|\theta\in\mathcal{I}\}$. The FS atomic norm is then 
	\begin{equation}
	\|\tilde{\mathbf{h}}\|_{\mathcal{A}_{\mathcal{I}}}=\inf_{\substack{\mathbf{a}(N_t,\theta_l)\in\mathcal{A_{\mathcal{I}}}\\\alpha_l\in\mathbb{C}}}\left\lbrace \sum_{l=1}^{L}|\alpha_l|:\tilde{\mathbf{h}}=\sum_{l=1}^{L}\alpha_l^*\mathbf{a}(N_t,\theta_l)\right\rbrace .\label{eq:10}
	\end{equation}
	
	Note that \eqref{eq:10} is equivalent to the following semi-definite program (SDP) \cite{c35}
	\begin{equation}
	\left\{ \begin{aligned}
	\|\tilde{\mathbf{h}}\|_{\mathcal{A}_{\mathcal{I}}}&=\mathop{\inf}_{\begin{subarray}{c}\mathbf{t}\in\mathbb{C}^{(2N_t-1)\times 1}\\ t\in\mathbb{R}\end{subarray}} \frac{1}{2N_t}\mathrm{Tr}(\mathrm{Toep}(\mathbf{t}))+\frac{t}{2}
	\\s.t.&~~~\left[ \begin{matrix}\mathrm{Toep}(\mathbf{t}) & \tilde{\mathbf{h}}\\
	\tilde{\mathbf{h}}^H & t\end{matrix}\right] \succeq 0,\mathbf{T}_{\beta}\succeq 0,
	\end{aligned}\right.\label{eq:fs_1d_sdp}
	\end{equation}
	where $\mathrm{Tr}(\cdot)$ denotes the trace, $\succeq 0$ indicates a semidefinite matrix, $t=\sum\nolimits_{l=1}^{L}|\alpha_l|$, and $\mathbf{T}_{\beta}$ is defined in \eqref{eq:tb}.
	
	According to \eqref{eq:y_1d}, the 1D channel estimation can be formulated as the following optimization problem:
	\begin{equation}
	\hat{\mathbf{h}}=\arg\min_{\tilde{\mathbf{h}}\in\mathbb{C}^{N_t\times 1}}\frac{1}{2}\|\tilde{\mathbf{y}}-\mathbf{X}^H\mathbf{F}^H\tilde{\mathbf{h}}\|_2^2+\mu\|\tilde{\mathbf{h}}\|_{\mathcal{A}_{\mathcal{I}}},\label{eq:ce_1d}
	\end{equation}
	where $\mu>0$ is the weight factor. In practice, we set $\mu \simeq \sigma_n\sqrt{N_t\log(N_t)}$.


The problem in \eqref{eq:ce_1d} has $n=\mathcal{O}(N_t)$ free variables and $m=2$ linear matrix inequations (LMI), and the $i$-th LMI has size of $k_i\times k_i$ with $k_i=\mathcal{O}(N_t)$. It follows from \cite{c38} that a primal-dual algorithm for \eqref{eq:ce_1d} has a computational complexity on the order of
\begin{equation}
\left(1+\sum_{i=1}^mk_i\right)^{\frac{1}{2}}n\left(n^2+n\sum_{i=1}^mk_i^2+\sum_{i=1}^mk_i^3\right)=\mathcal{O}(N_t^{4.5}).\label{eq:comp}
\end{equation}
By arguments similar to those above, the original atomic norm method \cite{c36} in the absence of prior knowledge has the same computational complexity of $\mathcal{O}(N_t^{4.5})$.

\subsection{Multiple Rx Antennas}
For the case of $N_t,N_r>1$, $\mathbf{Y}$ in \eqref{eq:Y} is vectorized as
\begin{equation}
\tilde{\mathbf{y}}=\mathrm{vec}(\mathbf{Y})=(\mathbf{X}^T\mathbf{F}^T\otimes\mathbf{I})\tilde{\mathbf{h}}+\tilde{\mathbf{n}},\label{eq:vec_y}
\end{equation} 
where $\mathbf{I}$ is the identity matrix of size $N_r$, $\tilde{\mathbf{h}}=\mathrm{vec}(\mathbf{H})=\sum\nolimits_{l=1}^{L}\alpha_l\mathbf{a}^*(N_t,\theta_l)\otimes\mathbf{a}(N_r,\phi_l)$ and $\tilde{\mathbf{n}}=\mathrm{vec}(\mathbf{N})$, with $\otimes$ being the Kronecker product.

Before solving the problem, we first extend the FS atomic norm in \cite{c33} to the 2D case as follows.

For the 2D case, we define $\mathcal{I}_1=(f_{L1},f_{H1})\subset[-\frac{1}{2},\frac{1}{2}]$, $\mathcal{I}_2=(f_{L2},f_{H2})\subset[-\frac{1}{2},\frac{1}{2}]$ and a 2-level Toeplitz matrix $\mathcal{T}(\mathbf{V})\in\mathbb{C}^{N_tN_r\times N_tN_r}$ formed by the elements of $\mathbf{V}$, where $\mathbf{V}$ is defined as $\mathbf{V}=[\mathbf{v}_{-N_r+1},\mathbf{v}_{-N_r+2},...,\mathbf{v}_{N_r-1}]$, with $\mathbf{v}_j=[v_j(-N_t+1),v_j(-N_t+2),...,v_j(N_t-1)]^T\in\mathbb{C}^{(2N_t-1)\times 1}$, $j=-N_r+1,-N_r+2,...,N_r-1$. More specifically, $\mathcal{T}(\mathbf{V})$ is in the form of
\begin{equation}
\begin{split}
&[\mathcal{T}(\mathbf{V})]_{pq}=\mathrm{Toep}(\mathbf{v}_{p-q}),\\
&[\mathrm{Toep}(\mathbf{v}_{j})]_{mn}=v_j(m-n),
\end{split}\label{eq:2toep}
\end{equation}
where $\mathrm{Toep}(\cdot)$ denotes the Toeplitz matrix whose first column is the last $N_t$ elements of the input vector, $j=-N_r+1,-N_r+2,...,N_r-1$, with $1\leq p,q\leq N_r$ denoting the block indices and $1\leq m,n\leq N_t$ denoting the element indices. Similar to \eqref{eq:beta}, we can write $\beta_1(f)$ and $\beta_2(f)$ according to $\mathcal{I}_1$ and $\mathcal{I}_2$, whose parameters are denoted as $\{r_{1,j}\}_{j=-1}^1$ and $\{r_{2,j}\}_{j=-1}^1$, respectively. Then the corresponding 2-level Toeplitz matrices, i.e., $\mathcal{T}_{\beta 1}\in\mathbb{C}^{(N_t-1)N_r\times (N_t-1)N_r}$ and $\mathcal{T}_{\beta 2}\in\mathbb{C}^{N_t(N_r-1)\times N_t(N_r-1)}$, are given by
\begin{equation}
[[\mathcal{T}_{\beta 1}]_{pq}]_{mn}=\sum\nolimits_{j=-1}^1 r_{1,j}v_{p-q}(m-n-j),\label{eq:fst1}
\end{equation}
where $1\leq p,q\leq N_r$, $1\leq m,n\leq N_t-1$, and
\begin{equation}
[\mathcal{T}_{\beta 2}]_{pq}=\sum\nolimits_{j=-1}^1 r_{2,j}\mathrm{Toep}(\mathbf{v}_{p-q-j}),\label{eq:fst2}
\end{equation}
where $1\leq p,q\leq N_r-1$.

Define the 2D FS atom as $\mathbf{b}(\theta,\phi)=\mathbf{a}(N_t,\theta)\otimes \mathbf{a}(N_r,\phi)\in\mathbb{C}^{N_tN_r}$, and the set of 2D FS atoms as $\mathcal{A}_{\mathcal{I}_1,\mathcal{I}_2}=\{\mathbf{b}(\theta,\phi)|\theta\in\mathcal{I}_1,\phi\in\mathcal{I}_2\}$. Then the 2D FS atomic norm of any signal $\mathbf{p}$ with respect to $\mathcal{A}_{\mathcal{I}_1,\mathcal{I}_2}$ is defined as
\begin{equation}
\|\mathbf{p}\|_{\mathcal{A}_{\mathcal{I}_1,\mathcal{I}_2}}=\inf_{\begin{subarray}{c}\alpha_k\in\mathbb{C} \\ \theta_k\in\mathcal{I}_1,\phi_k\in\mathcal{I}_2\end{subarray}}\{\sum_{k=1}^K|\alpha_k|:\mathbf{p}=\sum_{k=1}^K\alpha_k\mathbf{b}(\theta_k,\phi_k)\}.
\end{equation}

\textbf{Lemma 1} \textit{Given $\mathcal{I}_1,\mathcal{I}_2\subset[-\frac{1}{2},\frac{1}{2}]$, a 2-level Toeplitz matrix $\mathcal{T}(\mathbf{V})\in\mathbb{C}^{N_tN_r\times N_tN_r}$ with $r=\mathrm{rank}(\mathcal{T}(\mathbf{V})) <\min (N_t,N_r)$ admits a unique FS Vandermonde decomposition as $\mathcal{T}(\mathbf{V})=\sum\nolimits_{k=1}^rc_k\mathbf{b}(\theta_k,\phi_k)\mathbf{b}^H(\theta_k,\phi_k)$ with $\theta_k\in\mathcal{I}_1$ and $\phi_k\in\mathcal{I}_2$, if and only if 
	\begin{equation}\left\{\begin{split}
	&\mathcal{T}(\mathbf{V})\succeq 0\\
	&\mathcal{T}_{\beta 1}\succeq 0\\
	&\mathcal{T}_{\beta 2}\succeq 0
	\end{split}\right. ,\label{eq:constra}
	\end{equation}
	where $\mathcal{T}(\mathbf{V})$, $\mathcal{T}_{\beta 1}$ and $\mathcal{T}_{\beta 2}$ are defined in \eqref{eq:2toep}, \eqref{eq:fst1} and \eqref{eq:fst2} respectively.}

The proof of \textit{\textbf{lemma 1}} can be found in the appendix.

For the 2D FS atomic norm, we have the corresponding SDP formulation as follow:

\textbf{Lemma 2} \textit{It holds that
	\begin{equation}
	\begin{split}
	\|\mathbf{p}\|_{\mathcal{A}_{\mathcal{I}_1,\mathcal{I}_2}}&=\min_{\mathbf{V},t}\frac{1}{2N_tN_r}\mathrm{Tr}(\mathcal{T}(\mathbf{V}))+\frac{1}{2}t,\\
	s.t.&~~\left[ \begin{matrix} \mathcal{T}(\mathbf{V})&\mathbf{p}\\\mathbf{p}^H&t\end{matrix} \right]\succeq 0,\mathcal{T}_{\beta 1}\succeq 0,\mathcal{T}_{\beta 2}\succeq 0.
	\end{split}\label{eq:l2}
	\end{equation}
}

The proof of \textit{\textbf{lemma 2}} can be found in the appendix.

From \eqref{eq:vec_y}, the 2D FS atomic norm for $\tilde{\mathbf{h}}$ is then
\begin{equation}
\|\tilde{\mathbf{h}}\|_{\mathcal{A}_{\mathcal{I}_1,\mathcal{I}_2}}=\inf_{\begin{subarray}{c}\alpha_l\in\mathbb{C} \\ \theta_l\in\mathcal{I}_1\\\phi_l\in\mathcal{I}_2\end{subarray}}\{\sum_{l=1}^L|\alpha_l|:\tilde{\mathbf{h}}=\sum_{l=1}^L\alpha_l\mathbf{a}^*(N_t,\theta_l)\otimes\mathbf{a}(N_r,\phi_l)\}.
\end{equation}

According to \textit{\textbf{lemma 2}}, the corresponding SDP formulation of $\|\tilde{\mathbf{h}}\|_{\mathcal{A}_{\mathcal{I}_1,\mathcal{I}_2}}$ is given by
\begin{equation}
\left\{ \begin{aligned}
\|\tilde{\mathbf{h}}\|_{\mathcal{A}_{\mathcal{I}_1,\mathcal{I}_2}}&=\mathop{\inf}_{\begin{subarray}{c}\mathbf{V}\in\mathbb{C}^{(2N_t-1)\times (2N_r-1)}\\ t\in\mathbb{R}\end{subarray}} \frac{1}{2N_tN_r}\mathrm{Tr}(\mathcal{T}(\mathbf{V}))+\frac{t}{2}
\\s.t.&~~~\left[ \begin{matrix}\mathcal{T}(\mathbf{V}) & \tilde{\mathbf{h}}\\
\tilde{\mathbf{h}}^H & t\end{matrix}\right] \succeq 0,\mathcal{T}_{\beta 1}\succeq 0,\mathcal{T}_{\beta 2}\succeq 0,
\end{aligned}\right.\label{eq:anm_fs}
\end{equation}
where the $\mathcal{T}_{\beta 1}$ and $\mathcal{T}_{\beta 2}$ are defined by \eqref{eq:fst1} and \eqref{eq:fst2}, respectively. 

Based on \eqref{eq:vec_y}, the 2D channel estimation can be formulated as the following optimization problem:
\begin{equation}
\hat{\mathbf{h}}=\arg\min_{\tilde{\mathbf{h}}\in\mathbb{C}^{N_tN_r\times 1}}\frac{1}{2}\|\tilde{\mathbf{y}}-(\mathbf{X}^T\mathbf{F}^T\otimes\mathbf{I})\tilde{\mathbf{h}}\|_2^2+\mu\|\tilde{\mathbf{h}}\|_{\mathcal{A}_{\mathcal{I}_1,\mathcal{I}_2}},\label{eq:min_h}
\end{equation}
where $\mu>0$ is the weight factor. In practice, we set $\mu \simeq \sigma_n\sqrt{N_tN_r\log(N_tN_r)}$.

Finally the estimated channel matrix is given by  $\hat{\mathbf{H}}=\mathrm{vec}^{-1}(\hat{\mathbf{h}})$, with $\hat{\mathbf{h}}$ being the solution to \eqref{eq:min_h}.

The problem in \eqref{eq:min_h} has $n=\mathcal{O}(N_tN_r)$ free variables and $m=3$ LMIs, and the $i$-th LMI has size of $k_i\times k_i$ with $k_i=\mathcal{O}(N_tN_r)$. It follows from \eqref{eq:comp} that a primal-dual algorithm for \eqref{eq:min_h} has a computational complexity of $\mathcal{O}(N_t^{4.5}N_r^{4.5})$, which equals to that of the original 2D atomic norm method in \cite{c36}.

\section{SIMULATION RESULTS}
\begin{figure}[htbp]
	\centering
	\subfigure[NMSE of 128$\times$1 channel estimation]{
		\begin{minipage}[t]{0.5\linewidth}
			\centering
			\includegraphics[width=1.65in,height=1.2in]{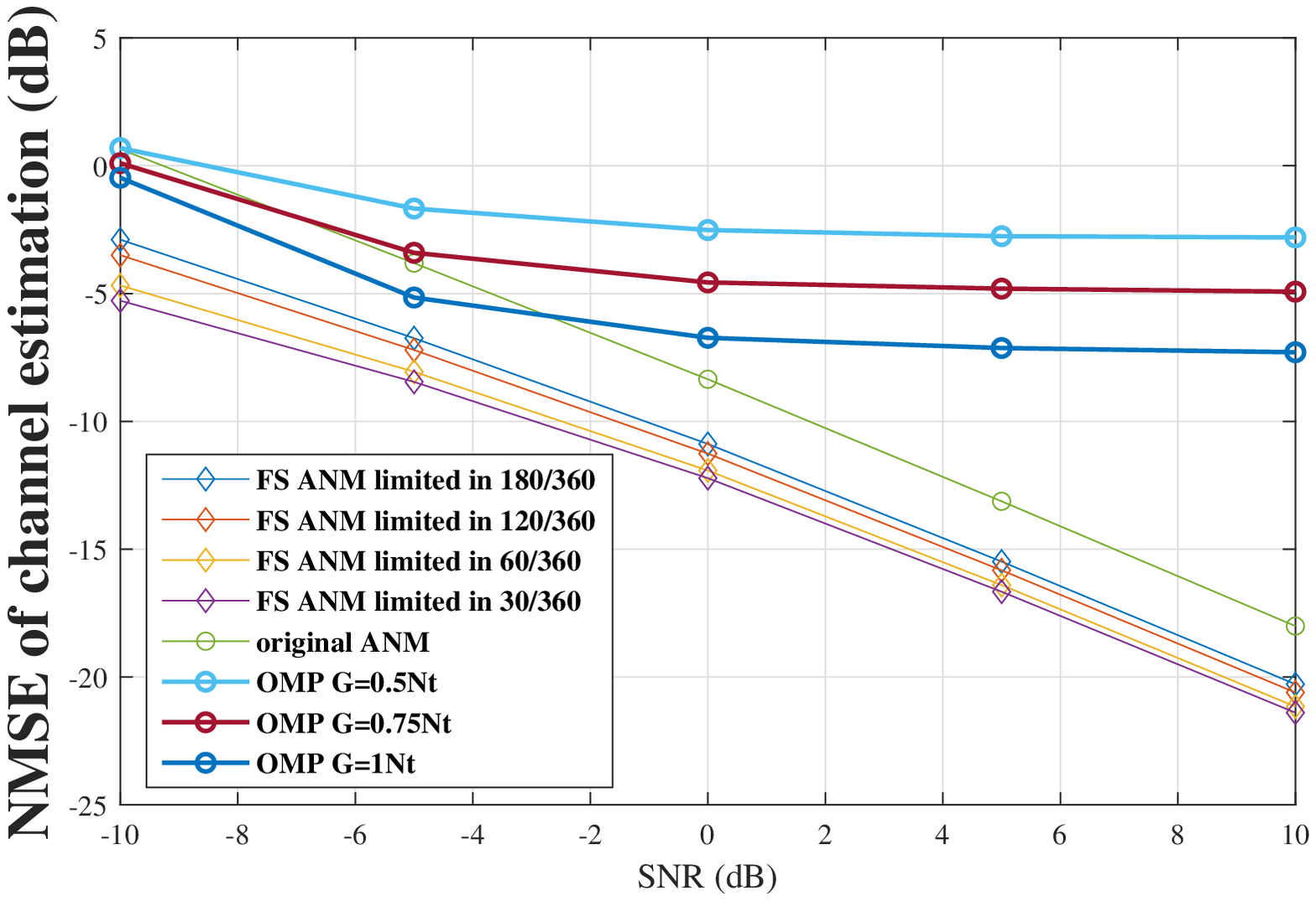}
			\label{fig:1d}
		\end{minipage}%
	}%
	\subfigure[NMSE of 16$\times$8 channel estimation]{
		\begin{minipage}[t]{0.5\linewidth}
			\centering
			\includegraphics[width=1.65in,height=1.25in]{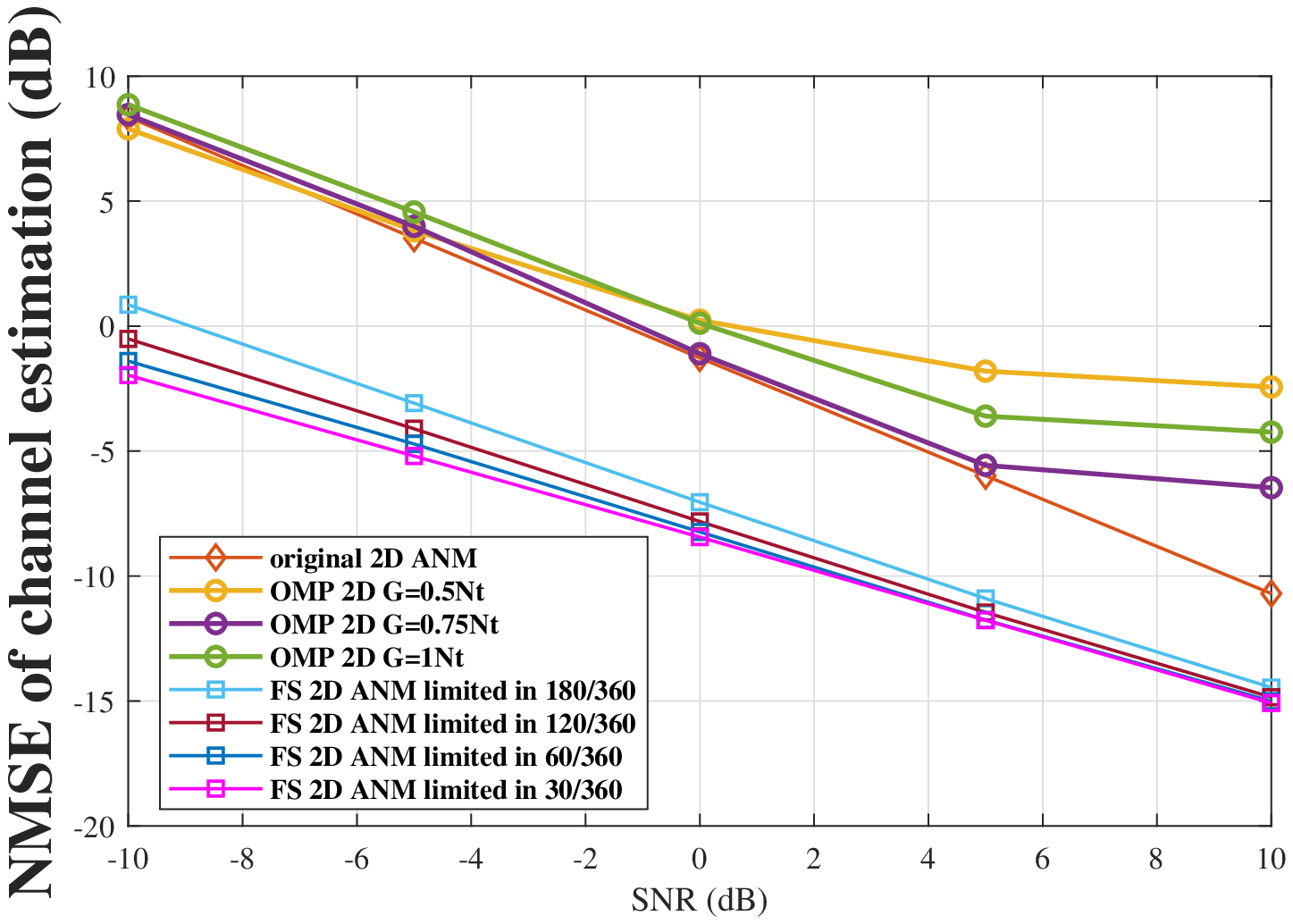}
			\label{fig:2d}
		\end{minipage}%
	}%
	\centering
	\caption{NMSE of channel estimation using the proposed algorithm: estimating (a) the 128$\times$1 ULA channel (b) the 16$\times$8 ULA channel, using the FS-ANM and OMP methods with different frequency constraints ($180^{\circ}$, $120^{\circ}$, $60^{\circ}$ and $30^{\circ}$) and different grids ($0.5N_t$, $0.75N_t$ and $N_t$) respectively.}
\end{figure}
In this section, we use simulation to illustrate the performance of the proposed algorithm. The pilot matrix is set as an identity matrix of size $S$. The wavelength of the signal is set as $\lambda=1\times 10^{-3}$m, i.e., the system is working on $0.3\mathrm{THz}$. Note that the scattering at THz frequencies induces more than 20 dB attenuation, which means that almost only the LoS component can be used for reliable high-rate transmission in THz communications. It is also worth pointing out that the performance loss induced by the consideration that only LoS component exists is negligible, since the number of NLoS components is quite limited and the power of NLoS components is much weaker (more than 20 dB) than that of LoS component in THz communications\cite{c37}. Therefore, the number of paths is set as $L=2$, with $\{\alpha_l\}_{l=1}^L$ generated by $\mathcal{CN}(0,1)$ and $\mathcal{CN}(0,0.1)$ respectively. The interval of the ULA is set as $d=\lambda/2$. For the 1D case, we set $N_t=128$, $N_r=1$ and $S=50$. $\{\phi_l\}_{l=1}^L$ is generated according to the distribution $\mathcal{U}(-\frac{1}{2},\frac{1}{2})$, with $\min_{i\ne j}|\phi_i-\phi_j|>1/N_t$. 
For the 2D case, we set $N_t=16$, $N_r=8$ and $S=16$. The frequencies $\{\theta_l\}_{l=1}^L$ and $\{\phi_l\}_{l=1}^L$ are generated by $\mathcal{U}(-\frac{1}{2},\frac{1}{2})$,
with $\min_{i\ne j}(|\phi_i-\phi_j|,|\theta_i-\theta_j|)>1/N_tN_r$. The frequency constraints are set as $\Omega\in\frac{2\pi }{360^{\circ}}\{180^{\circ},120^{\circ},60^{\circ},30^{\circ}\}$. The corresponding $\mathcal{I}_1=[\min\frac{1}{2}\sin(\tilde{\theta}),\max\frac{1}{2}\sin(\tilde{\theta})]$, $\mathcal{I}_2=[\min\frac{1}{2}\sin(\tilde{\phi}),\max\frac{1}{2}\sin(\tilde{\phi})]$, where $\tilde{\theta}\in[\bar{\theta}-\frac{\Omega}{2},\bar{\theta}+\frac{\Omega}{2}]$, $\tilde{\phi}\in[\bar{\phi}-\frac{\Omega}{2},\bar{\phi}+\frac{\Omega}{2}]$.

For the first experiment, as shown in Fig. \ref{fig:1d}, to compare with the FS-ANM algorithm, we consider two existing methods, i.e., the OMP method and the original ANM algorithm in \cite{c36}. The result indicates that the proposed FS-ANM always outperforms the mentioned two existing method with a gap of about $5\mathrm{dB}$. As an off-grid algorithm, the original ANM shows a higher performance than that of the on-grid OMP method, which is limited by the density of grid setting. However, the original ANM performs still worse than the proposed FS-ANM. For the second experiment, as shown in Fig. \ref{fig:2d}, we also consider the OMP method and the original ANM in \cite{c36} as the reference. It can be seen that the precision of FS-ANM is generally higher than the original ANM, with the gap of about $10\mathrm{dB}$ and $5\mathrm{dB}$ at $\mathrm{SNR}=-10\mathrm{dB}$ and $\mathrm{SNR}=10\mathrm{dB}$ respectively. 
On the other hand, prior knowledge is more informed, i.e., the frequency range becomes smaller, the FS-ANM algorithm achieves higher accuracy. 

\section{CONCLUSIONS}
In this paper, we have proposed a new off-grid MIMO channel estimation method for mmWave/THz systems that can exploit the prior knowledge on the ranges of AoD/AoA, based on frequency-selective atomic norm minimization. Simulation results indicates that the proposed algorithm significantly outperforms the existing on-grid/off-grid channel estimators.

\begin{appendix} 

\textit{Proof of Lemma 1.} We first show the ``if" part. It follows from $\mathcal{T}(\mathbf{V})\succeq 0$ that $\mathcal{T}(\mathbf{V})$ admits a unique Vandermonde decomposition\cite{c33}. Therefore, it suffices to show $\theta_k\in\mathcal{I}_1$ and $\phi_k\in\mathcal{I}_2$ under the condition $\mathcal{T}_{\beta 1}\succeq 0$ and $\mathcal{T}_{\beta 2}\succeq 0$. According to \eqref{eq:2toep}, the element of $\mathcal{T}(\mathbf{V})$ is given by
\begin{equation}\begin{split}
[[\mathcal{T}(\mathbf{V})]_{pq}]_{mn}=&v_{p-q}(m-n)\\
=&\sum\nolimits_{k=1}^rc_ke^{i2\pi(m-n)\theta_k}e^{i2\pi(p-q)\phi_k}.\end{split}
\end{equation}
Then we have
\begin{equation}\begin{split}
[[\mathcal{T}_{\beta 1}]_{pq}]_{mn}=&\sum_{j=-1}^1 r_{1,j}v_{p-q}(m-n-j)\\
=&\sum_{j=-1}^1 r_{1,j} \sum_{k=1}^rc_ke^{i2\pi(m-n-j)\theta_k}e^{i2\pi(p-q)\phi_k}\\
=&\sum_{k=1}^rc_ke^{i2\pi(m-n)\theta_k}e^{i2\pi(p-q)\phi_k}\underbrace{\sum_{j=-1}^1 r_{1,j}e^{-i2\pi j\theta_k}}_{\beta_1(\theta_k)}\\
=&\sum_{k=1}^rc_k\beta_1(\theta_k)e^{i2\pi(m-n)\theta_k}e^{i2\pi(p-q)\phi_k},
\end{split}\end{equation}
and hence
\begin{equation}\begin{split}
\mathcal{T}_{\beta 1}=&\sum\nolimits_{k=1}^rc_k\beta_1(\theta_k)\mathbf{b}_1(\theta_k,\phi_k)\mathbf{b}_1(\theta_k,\phi_k)^H\\
=&\mathbf{B}_1\mathrm{diag}(c_1\beta_1(\theta_1),...,c_r\beta_1(\theta_r))\mathbf{B}_1^H,\label{eq:b1}
\end{split}\end{equation}
where $\mathbf{b}_1(\theta_k,\phi_k)=\mathbf{a}(N_t-1,\theta_k)\otimes \mathbf{a}(N_r,\phi_k)$, $\mathbf{B}_1=[\mathbf{b}_1(\theta_1,\phi_1),...,\mathbf{b}_1(\theta_r,\phi_r)]$. 

According to \eqref{eq:fst2} and \eqref{eq:2toep}, we have
\begin{equation}
[[\mathcal{T}_{\beta 2}]_{pq}]_{mn}=\sum\nolimits_{j=-1}^1 r_{2,j}v_{p-q-j}(m-n).
\end{equation}
Similarly, we can get
\begin{equation}
\mathcal{T}_{\beta 2}
=\mathbf{B}_2\mathrm{diag}(c_1\beta_2(\phi_1),...,c_r\beta_2(\phi_r))\mathbf{B}_2^H,\label{eq:b2}
\end{equation}
where $\mathbf{B}_2=[\mathbf{b}_2(\theta_1,\phi_1),...,\mathbf{b}_2(\theta_r,\phi_r)]$, with $\mathbf{b}_2(\theta_k,\phi_k)=\mathbf{a}(N_t,\theta_k)\otimes \mathbf{a}(N_r-1,\phi_k)$.

Since $r\leq \min (N_t(N_r-1),(N_t-1)N_r)$, $\mathbf{B}_1$ and $\mathbf{B}_2$ have full column ranks. Using \eqref{eq:constra}, \eqref{eq:b1} and \eqref{eq:b2}, we have
\begin{equation}\left\lbrace \begin{split}
&\mathrm{diag}(c_1\beta_1(\theta_1),...,c_r\beta_1(\theta_r))=\mathbf{B}_1^\dag \mathcal{T}_{\beta 1} \mathbf{B}_1^{\dag H}\succeq 0,\\
&\mathrm{diag}(c_1\beta_2(\phi_1),...,c_r\beta_2(\phi_r))=\mathbf{B}_2^\dag \mathcal{T}_{\beta 2} \mathbf{B}_2^{\dag H}\succeq 0,
\end{split}\right.\end{equation}
where $\dag$ denotes the matrix pseudo-inverse operator. Thus $\forall k$, $c_k\beta_1(\theta_k)\geq0$, $c_k\beta_2(\phi_k)\geq0$. Since $c_k>0$, we have $\beta_1(\theta_k)\geq0$ and $\beta_2(\phi_k)\geq0$. By the property of $\beta(f)$, we finally have $\theta_k\in\mathcal{I}_1$, $\phi_k\in\mathcal{I}_2$, $k=1,...,r$.

The ``only if" part can be shown by similar arguments. Given $\mathcal{T}(\mathbf{V})=\sum\nolimits_{k=1}^rc_k\mathbf{b}(\theta_k,\phi_k)\mathbf{b}^H(\theta_k,\phi_k)$, it is evident that $\mathcal{T}(\mathbf{V})\succeq 0$. Then on the basis of \eqref{eq:b1}, \eqref{eq:b2} and the property of $\beta(f)$, we have $\mathcal{T}_{\beta 1}\succeq 0$ and $\mathcal{T}_{\beta 2}\succeq 0$.

\textit{Proof of Lemma 2.} Let $F^{\star}$ be the optimal objective value of \eqref{eq:l2}. We need to show that $\|\mathbf{p}\|_{\mathcal{A}_{\mathcal{I}_1,\mathcal{I}_2}}=F^{\star}$.

To begin with, we first show that $F^{\star}\leq \|\mathbf{p}\|_{\mathcal{A}_{\mathcal{I}_1,\mathcal{I}_2}}$. Let $\mathbf{p}=\sum_kc_k\mathbf{b}(\theta_k,\phi_k)\psi_k$ be an 2D FS Vandermonde decomposition of $\mathbf{p}$ on $\mathcal{I}_1$ and $\mathcal{I}_2$, with $|\psi_k|^2=1$. Then let $\mathbf{V}$ conform to $\mathcal{T}(\mathbf{V})=\sum\nolimits_kc_k\mathbf{b}(\theta_k,\phi_k)\mathbf{b}^H(\theta_k,\phi_k)$ and $t=\sum_kc_k$. By \textit{\textbf{lemma 1}}, we have $\mathcal{T}_{\beta 1}\succeq 0$ and $\mathcal{T}_{\beta 2}\succeq 0$. Furthermore, it holds that
\begin{equation}
\left[ \begin{matrix} \mathcal{T}(\mathbf{V})&\mathbf{p}\\\mathbf{p}^H&t\end{matrix} \right]=\sum_kc_k\left[ \begin{matrix} \mathbf{b}(\theta_k,\phi_k)\\\bar{\psi_k}\end{matrix} \right]\left[ \begin{matrix} \mathbf{b}(\theta_k,\phi_k)\\\bar{\psi_k}\end{matrix} \right]^H\succeq 0.
\end{equation}
Thus the constructed $t$ and $\mathbf{V}$ are a feasible solution to the problem \eqref{eq:l2}, with the objective value calculated as
\begin{equation}
\frac{1}{2N_tN_r}\mathrm{Tr}(\mathcal{T}(\mathbf{V}))+\frac{1}{2}t=\sum_kc_k.
\end{equation}
Therefore, it holds that $F^{\star}\leq \sum_kc_k$. Since the inequality holds for any FS atomic decomposition of $\mathbf{p}$ on $\mathcal{I}_1$ and $\mathcal{I}_2$, we have that $F^{\star}\leq \|\mathbf{p}\|_{\mathcal{A}_{\mathcal{I}_1,\mathcal{I}_2}}$ based on the definition of $\|\mathbf{p}\|_{\mathcal{A}_{\mathcal{I}_1,\mathcal{I}_2}}$.

Next we will show that $F^{\star}\geq \|\mathbf{p}\|_{\mathcal{A}_{\mathcal{I}_1,\mathcal{I}_2}}$. We suppose that $(t^{\star},\mathbf{V}^{\star})$ is the optimal solution to \eqref{eq:l2}. By the fact that $\mathcal{T}(\mathbf{V}^{\star})\succeq 0$, $\mathcal{T}_{\beta 1}^{\star}\succeq 0$ and $\mathcal{T}_{\beta 2}^{\star}\succeq 0$, according to \textit{\textbf{lemma 1}}, $\mathcal{T}(\mathbf{V}^{\star})$ has an FS Vandermonde decomposition on $\mathcal{I}_1$ and $\mathcal{I}_2$ given by 
\begin{equation}
\mathcal{T}(\mathbf{V}^{\star})=\sum\nolimits_{k=1}^{r^{\star}}c_k^{\star}\mathbf{b}(\theta_k^{\star},\phi_k^{\star})\mathbf{b}^H(\theta_k^{\star},\phi_k^{\star}).
\end{equation}
Since $\left[ \begin{matrix} \mathcal{T}(\mathbf{V}^{\star})&\mathbf{p}\\\mathbf{p}^H&t^{\star}\end{matrix} \right]\succeq 0$, $\mathbf{p}$ lies in the range space of $\mathcal{T}(\mathbf{V}^{\star})$ and thus has an FS atomic decomposition given by 
\begin{equation}
\mathbf{p}=\sum\nolimits_{k=1}^{r^{\star}}c_k^{\star}\mathbf{b}(\theta_k^{\star},\phi_k^{\star})\psi_k^{\star},~~ |\psi_k^{\star}|^2=1,\theta_k^{\star}\in\mathcal{I}_1,\phi_k^{\star}\in\mathcal{I}_2,
\end{equation}
which achieves the FS atomic norm. Furthermore, it holds that 
\begin{equation}
\begin{split} 
&t^{\star}\geq \mathbf{p}^H [\mathcal{T}(\mathbf{V}^{\star})]^{\dagger}\mathbf{p}=\sum\nolimits_{k=1}^{r^{\star}}c_k^{\star},\\
&\frac{1}{N_tN_r}\mathrm{Tr}(\mathcal{T}(\mathbf{V}^{\star}))=\sum\nolimits_{k=1}^{r^{\star}}c_k^{\star}.
\end{split}
\end{equation}
Thus we have
\begin{equation}
F^{\star}=\frac{1}{2N_tN_r}\mathrm{Tr}(\mathcal{T}(\mathbf{V}^{\star}))+\frac{t^{\star}}{2}\geq \sum\nolimits_{k}c_k^{\star}\geq \|\mathbf{p}\|_{\mathcal{A}_{\mathcal{I}_1,\mathcal{I}_2}}.
\end{equation} 
Since that $F^{\star}\leq \|\mathbf{p}\|_{\mathcal{A}_{\mathcal{I}_1,\mathcal{I}_2}}$ and $F^{\star}\geq \|\mathbf{p}\|_{\mathcal{A}_{\mathcal{I}_1,\mathcal{I}_2}}$ have both been shown, we conclude that $F^{\star}=\|\mathbf{p}\|_{\mathcal{A}_{\mathcal{I}_1,\mathcal{I}_2}}$. Thus \textit{\textbf{lemma 2}} is proved.

\end{appendix}


\begin{thebibliography}{00}
\bibitem{c01} P. Wang, Y. Li, L. Song, B. Vucetic, ``Multi-gigabit millimeter wave wireless communications for 5G: From fixed access to cellular networks", \textit{IEEE Commun. Mag.}, vol. 53, no. 1, pp. 168-178, Jan. 2015.
\bibitem{c02} V. W. Wong, R. Schober, D. W. K. Ng, L.-C. Wang, Key Technologies for 5G Wireless Systems, Cambridge, U.K.:Cambridge Univ. Press, 2017.
\bibitem{c13} E. A. Omar, S. Rajagopal, S. Abu-Surra, Z. Pi, R. W. Heath, Jr., ``Spatially sparse precoding in millimeter wave MIMO systems", \textit{IEEE Trans. Wirel. Commun.}, vol. 13, no. 3, pp. 1499-1513, Mar. 2014.
\bibitem{c14} L. Zhao, G. Geraci, T. Yang, D. W. K. Ng, J. Yuan, ``A tone-based AoA estimation and multiuser precoding for millimeter wave massive MIMO", \textit{IEEE Trans. Commun.}, vol. 65, no. 12, pp. 5209-5225, Dec. 2017.
\bibitem{c15} J. Wang, Z. Lan, C. Pyo, T. Baykas, C. S. Sum, M. A. Rahman, R. F. F. Kojima, I. L. H. Harada, S. Kato, ``Beam codebook based beamforming protocol for multi-Gbps millimeter-wave WPAN systems", \textit{IEEE J. Sel. Areas Commun.}, vol. 27, no. 8, pp. 1390-1399, Oct. 2009.
\bibitem{c16} A. Alkhateeb, O. El Ayach, G. Leus, R. W. Heath, ``Channel estimation and hybrid precoding for millimeter wave cellular systems", \textit{IEEE J. Sel. Top. Signal Process.}, vol. 8, no. 5, pp. 831-846, Oct. 2014.
\bibitem{c17} T. S. Rappaport, R. W. Heath, R. C. Daniels, J. N. Murdock, Millimeter Wave Wireless Communications, Englewood Cliffs, NJ, USA:Prentice-Hall, 2015.
\bibitem{c18} Z. Guo, X. Wang, W. Heng, ``Millimeter-wave channel estimation based on two-dimensional beamspace MUSIC method", \textit{IEEE Trans. Wirel. Commun.}, vol. 16, no. 8, pp. 5384-5394, Aug. 2017.
\bibitem{c19} X. Rao, V. K. N. Lau, ``Distributed compressive CSIT estimation and feedback for FDD multi-user massive MIMO systems", \textit{IEEE Trans. Signal Process.}, vol. 62, no. 12, pp. 3261-3271, Jun. 2014.
\bibitem{c20} W. Ding, F. Yang, W. Dai, J. Song, ``Time-frequency joint sparse channel estimation for MIMO-OFDM systems", \textit{IEEE Commun. Lett.}, vol. 19, no. 1, pp. 58-61, Jan. 2015.
\bibitem{c21} W. U. Bajwa, J. Haupt, A. M. Sayeed, R. Nowak,``Compressed channel sensing: A new approach to estimating sparse multipath channels", \textit{Proc. IEEE}, vol. 98, no. 6, pp. 1058-1076, Jun. 2010.

\bibitem{c22} A. M. Sayeed, ``Deconstructing multiantenna fading channels", \textit{IEEE Trans. Signal Process.}, vol. 50, no. 10, pp. 2563-2579, Oct. 2002.

\bibitem{c23} J. Lee, G. T. Gil, Y. H. Lee, ``Channel estimation via orthogonal matching pursuit for hybrid MIMO systems in millimeter wave communications", \textit{IEEE Trans. Commun.}, vol. 64, no. 6, pp. 2370-2386, Jun. 2016.

\bibitem{c24} Z. Marzi, D. Ramasamy, U. Madhow, ``Compressive channel estimation and tracking for large arrays in mm-wave picocells", \textit{IEEE J. Sel. Top. Signal Process.}, vol. 10, no. 3, pp. 514-527, Apr. 2016.

\bibitem{c25} C. Hu, L. Dai, T. Mir, Z. Gao, J. Fang, ``Super-resolution channel estimation for mmWave massive MIMO with hybrid precoding", \textit{IEEE Trans. Veh. Technol.}, vol. 67, no. 9, pp. 8954-8958, Sep. 2018.

\bibitem{c26} S. Pejoski, V. Kafedziski, ``Estimation of sparse time dispersive channels in pilot aided OFDM using atomic norm", \textit{IEEE Commun. Lett.}, vol. 4, no. 4, pp. 397-400, Aug. 2015.



\bibitem{c29} Y. Zhang, G. Zhang, X. Wang, ``Array covariance matrix-based atomic norm minimization for off-grid coherent direction-of-arrival estimation", \textit{Proc. IEEE Int. Conf. Acoust. Speech Signal Process. (ICASSP)}, pp. 3196-3200, Mar. 2017.

\bibitem{c30} Z. Tian, Z. Zhang, Y. Wang, ``Low-complexity optimization for two-dimensional direction-of-arrival estimation via decoupled atomic norm minimization", in \textit{Proc. IEEE Int. Conf. Acoust. Speech Signal Process. (ICASSP)}, pp. 3071-3075, Mar. 2017.


\bibitem{c32} A. Alkhateeb, O. E. Ayach, G. Leus, R. W. Heath, ``Channel estimation and hybrid precoding for millimeter wave cellular systems", \textit{IEEE J. Sel. Top. Signal Process.}, vol. 8, no. 5, pp. 831-846, Jul. 2014.

\bibitem{c33} Z. Yang, L. Xie, ``Frequency-selective Vandermonde decomposition of Toeplitz matrices with applications", \textit{Signal Process.}, vol. 142, pp. 157-167, Jan. 2018.

\bibitem{c34} D. Yang, L. -L. Yang, L. Hanzo, ``DFT-based beamforming weight-vector codebook design for spatially correlated channels in the unitary precoding aided multiuser downlink", \textit{Proc. IEEE Int. Conf. Commun. (ICC)}, pp. 1-5, May 2010.

\bibitem{c35} G. Tang, B. N. Bhaskar, P. Shah, B. Recht, ``Compressed sensing off the grid", \textit{IEEE Trans. Inf. Theory}, vol. 59, no. 11, pp. 7465-7490, Nov. 2013.

\bibitem{c36} Z. Yang, L. Xie, P. Stoica, ``Vandermonde decomposition of multilevel Toeplitz matrices with application to multidimensional super-resolution", \textit{IEEE Trans. Inf. Theory}, vol. 62, no. 6, pp. 3685-3701, Jun. 2016.

\bibitem{c37} X. Gao, L. Dai, Y. Zhang, T. Xie, X. Dai, Z. Wang, ``Fast channel tracking for terahertz beamspace massive MIMO systems", \textit{IEEE Trans. Veh. Technol.}, vol. 66, no. 7, pp. 5689-5696, Jul. 2017.

\bibitem{c38} A. Ben-Tal, A. K. Nemirovski, Lectures on modern convex optimization: Analysis algorithms and engineering applications, PA, Philadelphia:SIAM, 2001.
\end{thebibliography}
\end{document}